\documentclass[twoside,twocolumn,english]{revtex4-1}
\usepackage[LGR,T1]{fontenc}
\usepackage[latin9]{inputenc}
\usepackage{geometry}
\usepackage{amsmath}
\usepackage{graphicx}

\makeatletter

\DeclareRobustCommand{\greektext}{%
  \fontencoding{LGR}\selectfont\def\encodingdefault{LGR}}
\DeclareRobustCommand{\textgreek}[1]{\leavevmode{\greektext #1}}
\ProvideTextCommand{\~}{LGR}[1]{\char126#1}

\RequirePackage{colortbl, tabularx}
\@ifundefined{comment}{}
  {
   
  }%

\makeatother

\usepackage{babel}
\begin{document}
\global\long\def\l{\lambda}
\global\long\def\ints{\mathbb{Z}}
\global\long\def\nat{\mathbb{N}}
\global\long\def\re{\mathbb{R}}
\global\long\def\com{\mathbb{C}}
\global\long\def\dff{\triangleq}
\global\long\def\df{\coloneqq}
\global\long\def\del{\nabla}
\global\long\def\cross{\times}
\global\long\def\der#1#2{\frac{d#1}{d#2}}
\global\long\def\bra#1{\left\langle #1\right|}
\global\long\def\ket#1{\left|#1\right\rangle }
\global\long\def\braket#1#2{\left\langle #1|#2\right\rangle }
\global\long\def\ketbra#1#2{\left|#1\right\rangle \left\langle #2\right|}
\global\long\def\paulix{\begin{pmatrix}0  &  1\\
 1  &  0 
\end{pmatrix}}
\global\long\def\pauliy{\begin{pmatrix}0  &  -i\\
 i  &  0 
\end{pmatrix}}
\global\long\def\sinc{\mbox{sinc}}
\global\long\def\ft{\mathcal{F}}
\global\long\def\dg{\dagger}
\global\long\def\bs#1{\boldsymbol{#1}}
\global\long\def\norm#1{\left\Vert #1\right\Vert }
\global\long\def\H{\mathcal{H}}
\global\long\def\tens{\varotimes}
\global\long\def\rationals{\mathbb{Q}}
 \global\long\def\tri{\triangle}
\global\long\def\lap{\triangle}
\global\long\def\e{\varepsilon}
\global\long\def\broket#1#2#3{\bra{#1}#2\ket{#3}}
\global\long\def\dv{\del\cdot}
\global\long\def\eps{\epsilon}
\global\long\def\rot{\vec{\del}\cross}
\global\long\def\pd#1#2{\frac{\partial#1}{\partial#2}}
\global\long\def\L{\mathcal{L}}
\global\long\def\inf{\infty}
\global\long\def\d{\delta}
\global\long\def\I{\mathbb{I}}
\global\long\def\D{\Delta}
\global\long\def\r{\rho}
\global\long\def\hb{\hbar}
\global\long\def\s{\sigma}
\global\long\def\t{\tau}
\global\long\def\O{\Omega}
\global\long\def\a{\alpha}
\global\long\def\b{\beta}
\global\long\def\th{\theta}
\global\long\def\l{\lambda}

\global\long\def\Z{\mathcal{Z}}
\global\long\def\z{\zeta}
\global\long\def\ord#1{\mathcal{O}\left(#1\right)}
\global\long\def\ua{\uparrow}
\global\long\def\da{\downarrow}
 \global\long\def\co#1{\left[#1\right)}
\global\long\def\oc#1{\left(#1\right]}
\global\long\def\tr{\mbox{tr}}
\global\long\def\o{\omega}
\global\long\def\nab{\del}
\global\long\def\p{\psi}
\global\long\def\pro{\propto}
\global\long\def\vf{\varphi}
\global\long\def\f{\phi}
\global\long\def\mark#1#2{\underset{#2}{\underbrace{#1}}}
\global\long\def\markup#1#2{\overset{#2}{\overbrace{#1}}}
\global\long\def\ra{\rightarrow}
\global\long\def\cd{\cdot}
\global\long\def\v#1{\vec{#1}}
\global\long\def\fd#1#2{\frac{\d#1}{\d#2}}
\global\long\def\P{\Psi}
\global\long\def\dem{\overset{\mbox{!}}{=}}
\global\long\def\Lam{\Lambda}
 \global\long\def\m{\mu}
\global\long\def\n{\nu}

\global\long\def\ul#1{\underline{#1}}
\global\long\def\at#1#2{\biggl|_{#1}^{#2}}
\global\long\def\lra{\leftrightarrow}
\global\long\def\var{\mbox{var}}
\global\long\def\E{\mathcal{E}}
\global\long\def\Op#1#2#3#4#5{#1_{#4#5}^{#2#3}}
\global\long\def\up#1#2{\overset{#2}{#1}}
\global\long\def\down#1#2{\underset{#2}{#1}}
\global\long\def\lb{\biggl[}
\global\long\def\rb{\biggl]}
\global\long\def\RG{\mathfrak{R}_{b}}
\global\long\def\g{\gamma}
\global\long\def\Ra{\Rightarrow}
\global\long\def\x{\xi}
\global\long\def\c{\chi}
\global\long\def\res{\mbox{Res}}
\global\long\def\dif{\mathbf{d}}
\global\long\def\dd{\mathbf{d}}
\global\long\def\grad{\vec{\del}}

\global\long\def\mat#1#2#3#4{\left(\begin{array}{cc}
 #1  &  #2\\
 #3  &  #4 
\end{array}\right)}
\global\long\def\col#1#2{\left(\begin{array}{c}
 #1\\
 #2 
\end{array}\right)}
\global\long\def\sl#1{\cancel{#1}}
\global\long\def\row#1#2{\left(\begin{array}{cc}
 #1  &  ,#2\end{array}\right)}
\global\long\def\roww#1#2#3{\left(\begin{array}{ccc}
 #1  &  ,#2  &  ,#3\end{array}\right)}
\global\long\def\rowww#1#2#3#4{\left(\begin{array}{cccc}
 #1  &  ,#2  &  ,#3  &  ,#4\end{array}\right)}
\global\long\def\matt#1#2#3#4#5#6#7#8#9{\left(\begin{array}{ccc}
 #1  &  #2  &  #3\\
 #4  &  #5  &  #6\\
 #7  &  #8  &  #9 
\end{array}\right)}
\global\long\def\su{\uparrow}
\global\long\def\sd{\downarrow}
\global\long\def\coll#1#2#3{\left(\begin{array}{c}
 #1\\
 #2\\
 #3 
\end{array}\right)}
\global\long\def\h#1{\hat{#1}}
\global\long\def\colll#1#2#3#4{\left(\begin{array}{c}
 #1\\
 #2\\
 #3\\
 #4 
\end{array}\right)}
\global\long\def\check{\checked}
\global\long\def\v#1{\vec{#1}}
\global\long\def\S{\Sigma}
\global\long\def\F{\Phi}
\global\long\def\M{\mathcal{M}}
\global\long\def\G{\Gamma}
\global\long\def\im{\mbox{Im}}
\global\long\def\til#1{\tilde{#1}}
\global\long\def\kb{k_{B}}
\global\long\def\k{\kappa}
\global\long\def\ph{\phi}
\global\long\def\el{\ell}
\global\long\def\en{\mathcal{N}}
\global\long\def\asy{\cong}
\global\long\def\sbl{\biggl[}
\global\long\def\sbr{\biggl]}
\global\long\def\cbl{\biggl\{}
\global\long\def\cbr{\biggl\}}
\global\long\def\hg#1#2{\mbox{ }_{#1}F_{#2}}
\global\long\def\J{\mathcal{J}}
\global\long\def\diag#1{\mbox{diag}\left[#1\right]}
\global\long\def\sign#1{\mbox{sgn}\left[#1\right]}
\global\long\def\T{\th}
\global\long\def\rp{\reals^{+}}

\title{L{\'e}vy Walks On Finite Intervals: A Step Beyond Asymptotics}

\author{Asaf Miron}

\address{Department of Physics of Complex Systems, The Weizmann Institute
of Science, Rehovot 7610001, Israel}
\begin{abstract}
A L{\'e}vy walk of order $\b$ is studied on an interval of length $L$,
driven out of equilibrium by different-density boundary baths. The
anomalous current generated under these settings is nonlocally related
to the density profile through an integral equation. While the asymptotic
solution to this equation is known, its finite-$L$ corrections remain
unstudied despite their importance in the study of anomalous transport.
Here a perturbative method for computing such corrections is presented
and explicitly demonstrated for the leading correction to the asymptotic
transport of a L{\'e}vy walk of order $\b=5/3$, which represents a broad
universal class of anomalous transport models. Surprisingly, many
other physical problems are described by similar integral equations,
to which the method introduced here can be directly applied.
\end{abstract}
\maketitle

\section{Introduction \label{sec:Introduction}}

The L{\'e}vy walk is a popular and well-studied model which describes
a variety of physical scenarios in which superdiffusive dynamics lead
to nonlocal stationary behavior. Nonlocality is manifested in the
mathematical description of the relevant observables in terms of an
integral equation with a power-law kernel \cite{mandelbrot1982fractal,drysdale1998levy,buldyrev2001average,lepri2011density,dhar2013exact,zaburdaev2015levy}. 

One fruitful application of the L{\'e}vy walk model is to the study of
anomalous heat transport in one-dimensional (1D) Hamiltonian systems
\cite{cipriani2005anomalous,zhao2006identifying,zoia2007fractional,lepri2011density,zaburdaev2011perturbation,dhar2013exact,liu2014anomalous,zaburdaev2015levy}.
When such systems are constrained to an interval of length $L$ and
driven out of equilibrium by heat baths of temperature difference
$\D T$, one observes an anomalous transport behavior in the asymptotic
large-$L$ limit: The energy current $J_{e}$ is found to scale as
$J_{e}\sim\D T/L^{1-\a}$ and the temperature profile is singular
at the boundaries. The ``anomalous exponent'' $\a\in\oc{0,1}$ gets
its name from the fact that Fourier's law predicts $\a=0$ \cite{grassberger2002heat,cipriani2005anomalous,lepri2011density,dhar2013exact}.
Active research of anomalous transport focuses on its universal features
in the asymptotic limit. The main questions include the classification
of models into different universality classes, precisely determining
the anomalous exponent $\a$ corresponding to each universality class
and the relation between $\a$ and the singular behavior of the accompanying
temperature profiles \cite{lepri2011density,lepri2016heat,cividini2017temperature}.

Although significant progress has recently provided several theoretical
predictions for the different universal values of $\a$ \cite{narayan2002anomalous,grassberger2002heat,wang2004intriguing,cipriani2005anomalous,lukkarinen2008anomalous,spohn2014nonlinear,popkov2015fibonacci},
obtaining conclusive experimental support is a difficult problem.
Namely, the asymptotic behavior predicted in theory for large-$L$
is hard to reach in numerical simulations and experiments due to finite-size
corrections \cite{cipriani2005anomalous}. In fact, it is generally
not known how large a system should be to ensure the asymptotic limit
is reached \cite{Miron2019}. Indeed, the literature contains numerous
observed values of $\a$ for a variety of models. These $\a$'s are
usually extracted from numerical simulations by fitting the observed
$L$ dependence of $J_{e}\left(L\right)$ to an inverse power law,
assuming that the system is safely inside the asymptotically large-$L$
regime. However, not all simulation results stand in agreement \cite{aoki2001fermi,grassberger2002heat,li2003anomalous,wang2004intriguing,cipriani2005anomalous,lepri2005studies,basile2007anomalous},
making it difficult to determine the different universality classes
and refute incompatible predictions. For this reason, insufficient
understanding of finite-size corrections poses a significant hurdle
which must be overcome to make progress.

Since L{\'e}vy particles are noninteracting, anomalous L{\'e}vy walk transport
is easier to study than that of Hamiltonian models. Imposing different-density
baths similarly gives rise to an anomalous walker current $J$ and
a corresponding singular density profile $P\left(x\right)$. This
setup was studied in Refs. \cite{lepri2011density,dhar2013exact}
where an integral equation relating the asymptotic current $J_{0}$
and density profile $P_{0}\left(x\right)$ was formulated and solved
exactly in Ref. \cite{dhar2013exact}. However, finite-size corrections
remain unstudied.

In this paper, a perturbative method is presented for computing finite-size
corrections to the asymptotic L{\'e}vy walk results in three steps: First,
the integral equation relating the asymptotic $J_{0}$ and $P_{0}\left(x\right)$
of Ref. \cite{dhar2013exact} is extended to include finite-size corrections.
Then, a perturbative method for computing the corrections order-by-order
in inverse powers of $L$ is introduced. Finally, the method is used
to explicitly compute the leading correction to the asymptotic $J_{0}$
and $P_{0}\left(x\right)$ for a L{\'e}vy walk of order $\b=5/3$, which
is expected to represent a broad universality class of anomalous transport
models with exponent $\a=1/3$ \cite{cipriani2005anomalous}. In this
case, the asymptotic current $J_{0}$ decays as $J_{0}\sim L^{-2/3}$,
whereas its leading correction $J_{1}$ is shown to decay as $J_{1}\sim L^{-1}$.
Thus, the asymptotic regime in which $J_{0}\gg J_{1}$ is reached
only when $L$ is very large, further illustrating the importance
of accounting for finite-size corrections. The intuitive explanation
behind the diffusive correction $J_{1}$ is that, although the width
of the L{\'e}vy walkers\textquoteright{} walk-time distribution diverges,
its mean is finite. Thus, although the walkers occasionally undergo
very long excursions, most of the walks last a small amount of time,
leading to diffusive transport. In an infinite system, the contribution
of finite walks vanishes. However, in finite systems they give rise
to corrections, the first of which is diffusive transport.

For the reasons noted above, these results constitute a crucial step
towards understanding finite-size corrections in anomalous transport
and, ultimately, in settling the debate on the different universality
classes and the precise values of the corresponding anomalous exponents.

A surprising corollary is that integral equations, which are similar
to the one derived for anomalous L{\'e}vy walk transport, also show up
in many additional physical scenarios. They appear, for example, in
the mean first-passage time of L{\'e}vy walkers on a finite interval with
absorbing boundaries \cite{buldyrev2001average}, in the anomalous
transport of a stochastic 1D gas model \cite{Miron2019}, in nonlocal
elasticity theory \cite{lazopoulos2006non,carpinteri2011fractional}
and in the viscosity of polymers in a solvent \cite{kirkwood1948intrinsic,douglas1989surface}.
As such, the method presented here for studying anomalous L{\'e}vy walk
transport can be directly applied to a diverse set of problems, spanning
across a wide range of research fields.

The paper is organized as follows: The L{\'e}vy walk model and nonequilibrium
setup are presented in Sec. \ref{sec:The-Model}. Section \ref{sec:A-Step-Beyond}
extends the asymptotic results of Ref. \cite{dhar2013exact} by first
deriving a more general integral equation, containing information
on both the asymptotic behavior and its corrections, and then presenting
a perturbative method for solving it. The method is explicitly used
to compute the leading correction to the known asymptotic behavior
for the L{\'e}vy walk of order $\b=5/3$ in Sec. \ref{sec:The-Leading-Correction}.
Section \ref{sec:Different beta} provides important details which
are relevant when applying the method to other values of $\b$ and
higher order corrections. Concluding remarks follow in Sec. \ref{sec:CONCLUSIONS}. 

\section{The Model \label{sec:The-Model}}

The L{\'e}vy walk model of order $\b$ describes particles moving at a
fixed velocity $v$ which evolve via random \textquotedblleft walks\textquotedblright{}
of duration $t$ drawn from the distribution 
\begin{equation}
\f\left(t\right)=\b t_{0}^{\b}\frac{\th\left[t-t_{0}\right]}{t^{\b+1}},\label{eq:phi(t)}
\end{equation}
where $1<\b<2$, $t_{0}$ is the minimal walk-time and $\th\left[\t\right]$
is the step function. All but the first moment of $\f\left(t\right)$
diverge, giving rise to rare, long walks that connect distant points
in the system \cite{grassberger2002heat,cipriani2005anomalous,lepri2011density,dhar2013exact}.
The model is studied on a 1D interval parameterized by $x\in\left[0,L\right]$.
Following Ref. \cite{dhar2013exact}, let $P\left(x,t\right)\dif x$
denote the number of walkers crossing the interval $\left(x,x+\dif x\right)$
at time $t$ and let $Q\left(x,t\right)\dif x\dif t$ denote the number
of walkers whose walk ends inside the interval $\left(x,x+\dif x\right)$
during the time interval $\left(t,t+\dif t\right)$. Correspondingly,
$P\left(x,t\right)$ is called the walker density and $Q\left(x,t\right)$
is called the turning-point density. It will prove useful to consider
the rescaled position $x\in\left[0,1\right]$, obtained by dividing
the position by $L$.

To model the nonequilibrium settings of anomalous transport, appropriate
boundary conditions must be imposed. Following Ref. \cite{dhar2013exact},
different density walker baths are imposed at the two ends of the
system by setting
\begin{equation}
Q\left(x\le0\right)=Q_{L}\text{ and }Q\left(x\ge1\right)=Q_{R}.\label{eq:baths}
\end{equation}
With these boundary conditions, the stationary walker current satisfies
the integral equation
\[
J_{exact}\left(x\right)=\frac{v}{2}\left(Q_{L}\int_{\frac{Lx}{v}}^{\infty}\dif\t\psi\left(\t\right)-Q_{R}\int_{\frac{L\left(1-x\right)}{v}}^{\infty}\dif\t\psi\left(\t\right)\right)
\]
\begin{equation}
+\frac{L}{2}\int_{0}^{1}\dif y\text{ }\sign{x-y}\p\left(\frac{L\left|x-y\right|}{v}\right)Q\left(y\right),\label{eq:Derrida's current}
\end{equation}
where $\psi\left(t\right)$, the probability of drawing a walk-time
larger than $t$, is given by 
\begin{equation}
\psi\left(t\right)=\int_{t}^{\infty}\dif\t\f\left(\t\right)=1+\th\left[t-t_{0}\right]\left(\left(\frac{t_{0}}{t}\right)^{\b}-1\right).\label{eq:psi beta}
\end{equation}
Eq. (\ref{eq:Derrida's current}) implies that the walker current
at position x is the sum of two contributions: The first line describes
the contribution coming from the two constant-density walker baths,
whereas the second line describes the contributions from walkers inside
the system. Since the system is in its steady state, the current must
be independent of $x$, i.e. $J_{exact}\left(x\right)\equiv J_{exact}$. 

It was also shown in Ref. \cite{dhar2013exact} that the turning point
density $Q\left(x\right)$ satisfies the self-consistent equation
\[
Q\left(x\right)=\frac{Q_{L}}{2}\psi\left(\frac{Lx}{v}\right)+\frac{Q_{R}}{2}\psi\left(\frac{L\left(1-x\right)}{v}\right)
\]
\begin{equation}
+\frac{L}{2v}\int_{0}^{1}\dif y\f\left(\frac{L\left|x-y\right|}{v}\right)Q\left(y\right),\label{eq:Q of x}
\end{equation}
and that the turning point density $Q\left(x\right)$ is related to
the walker density $P\left(x\right)$ by
\begin{equation}
P\left(x\right)=\frac{\b t_{0}}{\b-1}Q\left(x\right)+\ord{\e^{\b-1}},\label{eq:Q P relation}
\end{equation}
where $\e=t_{0}v/L$ plays the role of a dimensionless inverse system-size.
Eqs. (\ref{eq:Derrida's current}), (\ref{eq:Q of x}) and (\ref{eq:Q P relation})
constitute the starting point of this study.

\section{A Step Beyond Asymptotics \label{sec:A-Step-Beyond}}

Since the solution of Eq. (\ref{eq:Derrida's current}) for $J_{exact}$
is hard to compute, the first step is to expand the equation in small
$\e$
\begin{equation}
J\approx A\e P'\left(x\right)-B\e^{\b-1}\int_{0}^{1}\dif y\frac{P'\left(y\right)}{\left|x-y\right|^{\b-1}},\label{eq:Fredholm 2nd}
\end{equation}
where $A=\frac{v\left(\b-1\right)}{2\left(2-\b\right)}$, $B=\frac{v}{2\b}$,
$P'\left(x\right)$ denotes the derivative of $P\left(x\right)$.
Note that corrections of $\ord{\e^{2\left(\b-1\right)}}$ have been
neglected and will be addressed later in Sec. \ref{sec:Different beta}
(also see Appendix A). Equation (\ref{eq:Fredholm 2nd}) is derived
by substituting $\p\left(t\right)$ into Eq. (\ref{eq:Derrida's current})
for $J_{exact}$, expanding up to linear order in $\e$ and employing
the relation between $P\left(x\right)$ and $Q\left(x\right)$ of
Eq. (\ref{eq:Q P relation}). A similar equation for A = 0 was derived
in Ref. \cite{dhar2013exact}.

Before proceeding to solve Eq. (\ref{eq:Fredholm 2nd}), let us first
establish some useful notations. The rightmost term in Eq. (\ref{eq:Fredholm 2nd})
is intuitively called the ``nonlocal'' term since it depends on
the values of $P'\left(x\right)$ across the entire system. Naturally,
the term $A\e P'\left(x\right)$ is then referred to as the ``local''
term and $J$ is called the ``source'' term. The integral Eq. (\ref{eq:Fredholm 2nd})
for $J$ is a weakly singular Fredholm integral equation (WSFIE) of
the second kind \cite{kress1989linear,moiseiwitsch2011integral,zemyan2012classical}.
It is called weakly singular since the integral kernel diverges for
$y=x$ yet, since $0<\b-1<1$, the singularity is integrable. Last,
when the unknown function appears both under the integral sign and
outside the integral, the equation is of the \textquotedblleft second
kind\textquotedblright{} but if it appears only under the integral
sign, it is of the \textquotedblleft first kind\textquotedblright .
Note that an equation for $J$, containing both a local term $\sim\ord{\e^{-1}}$
and a non-local term $\sim\ord{\e^{\b-1}}$, is obtained whenever
the walk-time distribution $\f\left(t\right)$ contains a short walk-time
cutoff mechanism. 
\begin{figure}
\includegraphics[scale=0.67]{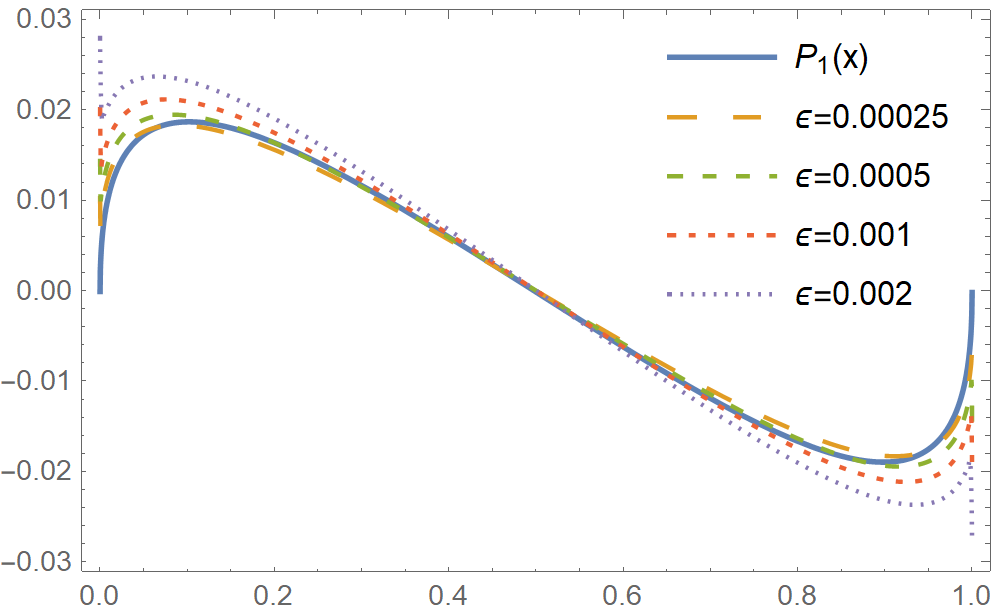}

\caption{A comparison between the density profile $P_{1}\left(x\right)$ of
Eq. (\ref{eq:P'1_1}) and the collapse of $\protect\e^{1/3}\left(P_{Num}\left(x\right)-P_{0}\left(x\right)\right)$
for different values of the inverse system size $\protect\e$. The
two are expected to identify as $\protect\e\protect\ra0$.\label{fig:P_1 collapse}}
\end{figure}

The simplest way to proceed is to take the asymptotic $\e\ra0$ limit
in Eq. (\ref{eq:Fredholm 2nd}). In this limit the local term vanishes
from Eq. (\ref{eq:Fredholm 2nd}), and with it all information about
finite-size corrections, reducing the equation to the WSFIE of the
first kind studied in Ref. \cite{dhar2013exact}. Although this WSFIE
can indeed be solved exactly by the Sonin formula \cite{samko1993fractional,buldyrev2001average},
the trade-off is that finite-size corrections remain out of reach.

Here, a method which preserves information about finite size corrections
is suggested instead. This method relies on the interplay between
the local term $\pro\e$ and the nonlocal term $\pro\e^{\b-1}$ to
construct an ansatz for $P'\left(x\right)$ and $J$ in the form of
a power-series in $\e^{\b-2}$, the ratio of the two scales, as
\begin{equation}
\begin{cases}
P'=P_{0}'+\e^{2-\b}P_{1}'+\e^{2\left(2-\b\right)}P_{2}'+...\\
J=\e^{\b-1}\left[\mathcal{J}_{0}+\e^{2-\b}\mathcal{J}_{1}+\e^{2\left(2-\b\right)}\mathcal{J}_{2}+...\right]
\end{cases},\label{eq:first few terms}
\end{equation}
where $P_{n}'\left(x\right)$ and $\mathcal{J}_{n}$ are independent
of $\e$. In turn, this allows replacing Eq. (\ref{eq:Fredholm 2nd})
by a hierarchy of WSFIEs of the first kind 
\begin{equation}
\begin{cases}
\mathcal{J}_{0}=-B\int_{0}^{1}\dif y\frac{P_{0}'\left(y\right)}{\left|x-y\right|^{\b-1}} & \text{at }\text{\ensuremath{\ord{\e^{\b-1}}}}\\
\mathcal{J}_{1}=AP_{0}'\left(x\right)-B\int_{0}^{1}\dif y\frac{P_{1}'\left(y\right)}{\left|x-y\right|^{\b-1}} & \text{at }\text{\ensuremath{\ord{\e}}}\\
\vdots\\
\mathcal{J}_{n}=AP_{n-1}'\left(x\right)-B\int_{0}^{1}\dif y\frac{P_{n}'\left(y\right)}{\left|x-y\right|^{\b-1}} & \text{at }\ord{\e^{\left(\b-1\right)+\left(2-\b\right)n}}
\end{cases}\label{eq:first few equations}
\end{equation}
where many of which can be solved using the Sonin formula \cite{samko1993fractional,buldyrev2001average}.
The first hierarchy equation, at $\ord{\e^{\b-1}}$, coincides with
the asymptotic equation of Ref. \cite{dhar2013exact} while the rest
provide increasingly higher-order, finite-size corrections which must
be solved in an iterative fashion. It is important to stress that
this method can be extended to additional WSFIEs of the second kind
which exhibit a similar interplay between the local and nonlocal terms,
even when the constant source term, e.g., $J$ in Eq. (\ref{eq:Fredholm 2nd}),
is replaced by a sufficiently well-behaved function of $x$ (see Appendix
B). In particular, it can be directly applied to the problems mention
in Sec. \ref{sec:Introduction} \cite{kirkwood1948intrinsic,douglas1989surface,lazopoulos2006non,carpinteri2011fractional,Miron2019}. 

\section{The Leading Correction For $\protect\b=5/3$ \label{sec:The-Leading-Correction}}

This method is next used to compute the leading correction to the
asymptotic density profile and current for a L{\'e}vy walk of order $\b=\frac{5}{3}$.
The generalization to different values of \textgreek{b} and higher-order
corrections is then discussed in Sec. \ref{sec:Different beta}.

Applying the ansatz
\begin{equation}
\begin{cases}
P'\left(x\right)=P_{0}'\left(x\right)+\e^{1/3}P_{1}'\left(x\right)+\ord{\e^{2/3}}\\
J=\e^{2/3}\mathcal{J}_{0}+\e\mathcal{J}_{1}+\ord{\e^{4/3}}
\end{cases},\label{eq:P' and J ansatz 5/3}
\end{equation}
to Eq. (\ref{eq:Fredholm 2nd}) for $J$ yields a hierarchy of WSFIEs
of the first kind. The first equation, appearing at $\ord{\e^{2/3}}$,
is simply the asymptotic equation studied in Ref. \cite{dhar2013exact}.
The solution is obtained by applying the Sonin formula \cite{samko1993fractional,buldyrev2001average}
(see Appendix B) and enforcing the boundary conditions in Eq. (\ref{eq:baths}).
One finds 
\begin{equation}
P_{0}'\left(x\right)=-\frac{b\mathcal{J}_{0}}{v\left(x\left(1-x\right)\right)^{1/6}}\text{ }\text{and}\text{ }\mathcal{J}_{0}=-\frac{av\D P}{b}\label{eq:J0 P0}
\end{equation}
 where $a=\G\left[\frac{5}{3}\right]/\G\left[\frac{5}{6}\right]^{2}$,
$b=\frac{5}{3\pi}$, $\G\left[x\right]$ is the gamma function and
$\D P\equiv P_{R}-P_{L}=\frac{5t_{0}}{2}\left(Q_{R}-Q_{L}\right)$
follows from Eq. (\ref{eq:Q P relation}).

To step beyond the known asymptotic results, let us consider the next
hierarchy equation for the leading correction, $P_{1}'\left(x\right)$.
This equation appears at $\ord{\e}$ and is given by

\begin{equation}
\frac{10}{3v}\left(vP_{0}'\left(x\right)-\mathcal{J}_{1}\right)=\int_{0}^{1}\dif y\frac{P_{1}'\left(y\right)}{\left|x-y\right|^{2/3}}.\label{eq:J_1 5/3}
\end{equation}
Due to the hierarchical structure of the ansatz of Eq. (\ref{eq:P' and J ansatz 5/3}),
$P_{0}'\left(x\right)$ enters this equation as a source term. Equation
(\ref{eq:J_1 5/3}) is also a WSFIE of the first kind since $P_{1}'\left(x\right)$
appears only inside the integral. Applying the Sonin formula \cite{samko1993fractional,buldyrev2001average}
yields 
\begin{equation}
P_{1}'\left(x\right)=-\frac{b}{v}\left(\frac{\mathcal{J}_{1}}{\left(x\left(1-x\right)\right)^{1/6}}+\frac{av\D P}{3^{1/2}2^{1/3}}I\left(x\right)\right),\label{eq:P'1_1}
\end{equation}
where $\e\mathcal{J}_{1}$ is the yet-unknown leading corrections
to the asymptotic current and $I\left(x\right)$ is given by
\begin{equation}
I\left(x\right)=\frac{1}{x^{\frac{1}{6}}}\der{}x\int_{x}^{1}\frac{\dif tt^{\frac{1}{3}}}{\left(t-x\right)^{\frac{1}{6}}}\der{}t\int_{0}^{t}\frac{\dif q\left(1-q\right)^{-\frac{1}{6}}}{q^{\frac{1}{3}}\left(t-q\right)^{\frac{1}{6}}}.\label{eq:I(x) formal}
\end{equation}
Manipulating $I\left(x\right)$ to its closed form requires careful
treatment since Eq. (\ref{eq:I(x) formal}) contains nontrivial improper
integrals. One finds
\[
I\left(x\right)=-\frac{2^{2/3}}{\left(x\left(1-x\right)\right)^{1/6}}-\frac{16x^{5/6}}{2^{1/3}5\left(1-x\right)^{7/6}}\biggl(G_{+}\left(x\right)
\]
\[
-\frac{5\left(2x+1\right)}{16x}G_{-}\left(x\right)\biggl)-\frac{\left(2x+1\right)\left(H_{+}\left(x\right)+H_{-}\left(x\right)\right)}{2\sqrt{x}\left(1-x\right)^{7/6}}
\]
\begin{equation}
+\frac{16\Gamma\left[\frac{5}{6}\right]\Gamma\left[\frac{8}{3}\right]\left(K_{+}\left(x\right)-K_{-}\left(x\right)\right)}{15\sqrt{\pi}\left(1-x\right)^{7/6}}\label{eq:I(x)}
\end{equation}
where $G_{\pm}\left(x\right)=F_{1}\left[\frac{7}{6}\pm\frac{1}{2};\frac{1}{6},\frac{1}{6};\frac{13}{6}\pm\frac{1}{2};\frac{2\sqrt{x}}{\sqrt{x}-1},\frac{2\sqrt{x}}{\sqrt{x}+1}\right]$,
$H_{\pm}\left(x\right)=\left(1\pm\sqrt{x}\right)^{2/3}F_{1}\left[\frac{2}{3};\frac{1}{6},\frac{1}{6};\frac{5}{3};\frac{\sqrt{x}\pm1}{\sqrt{x}\mp1},1\right]$
and $K_{\pm}\left(x\right)=\left(1\pm\sqrt{x}\right)^{5/3}\hg 21\left[\frac{1}{6},\frac{5}{3};\frac{5}{2};\frac{\sqrt{x}\pm1}{\sqrt{x}\mp1}\right]$.
Here $F_{1}\left[a;b_{1},b_{2};c;z_{1},z_{2}\right]$ is the Appell
hypergeometric function and $\hg 21\left[a;b;c;z\right]$ is the hypergeometric
function of the second kind. 

The function $I\left(x\right)$ has two interesting properties: First,
it is easy to show that the hierarchy equation (\ref{eq:J_1 5/3})
is symmetric under reflections $x\ra1-x$ and so $I\left(x\right)$
must too respect this symmetry. Induction can be used to extend this
argument to all hierarchy equations (see Appendix B). Second, one
can also show that, near the left boundary of the system, $I\left(x\right)$
behaves as 
\begin{equation}
I\left(x\ra0\right)\pro x^{-1/2}+\ord{x^{-1/6}}.\label{eq:boundary singularity}
\end{equation}
This implies that, for any finite $\e$, the boundary singularity
of the leading correction $P_{1}'\left(x\right)$ dominates over that
of the asymptotic solution $P_{0}'\left(x\right)$. 

Having found the closed-form solution for $P_{1}'\left(x\right)$,
the final step is to determine $\mathcal{J}_{1}$. This is done by
integrating Eq. (\ref{eq:P'1_1}) for $P_{1}\left(x\right)$ with
the appropriate boundary conditions. Since the asymptotic results
already use $P_{0}\left(1\right)-P_{0}\left(0\right)=\D P$ in Eq.
(\ref{eq:J0 P0}), the corrections must satisfy $P_{n}\left(0\right)=P_{n}\left(1\right)=0$
for all $n>0$. With these boundary conditions, $\mathcal{J}_{1}$
is given by 
\begin{equation}
\mathcal{J}_{1}=2^{4/3}3^{-1/2}av\D P.\label{eq:J1}
\end{equation}

\begin{figure}
\includegraphics[scale=0.65]{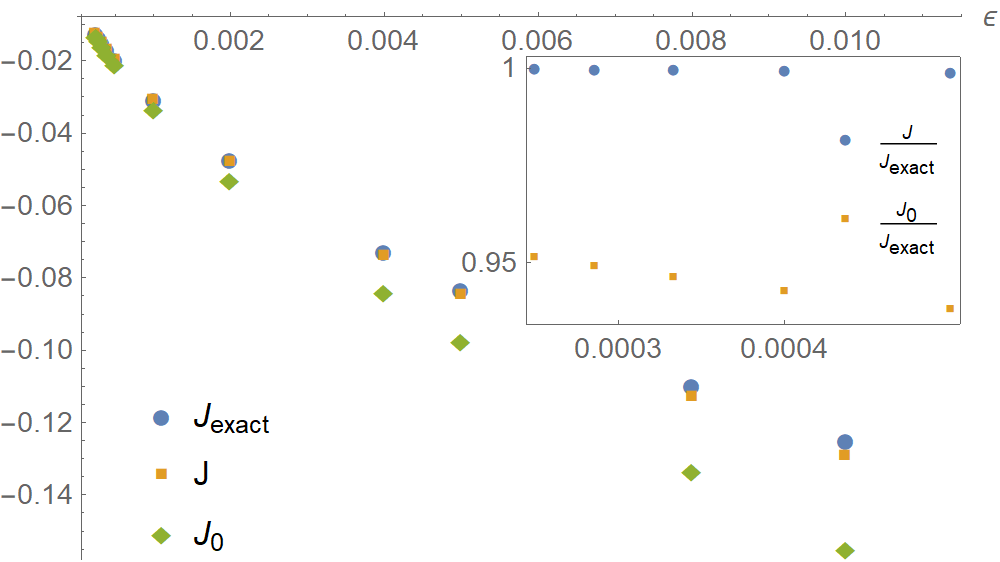}

\caption{The current $J$ predicted in Eq. (\ref{eq:P' and J}) (blue circles)
is compared to the asymptotic current $J_{0}$ of Eq. (\ref{eq:J0 P0})
(orange squares) and to the exact current $J_{exact}$ (green diamonds)
which is obtained by numerically solving $Q\left(x\right)$ of Eq.
(\ref{eq:Q of x}) and substituting the solution into Eq. (\ref{eq:Derrida's current}).
Inset: The ratios $\frac{J}{J_{exact}}$ (blue circles) and $\frac{J_{0}}{J_{exact}}$
(orange squares) are compared. The parameters are $v=t_{0}=1$ and
$\protect\D Q=1\protect\ra\protect\D P=5/2$. \label{fig:J}}
\end{figure}

Substituting $\mathcal{J}_{1}$ back inside Eq. (\ref{eq:P'1_1})
for $P_{1}'\left(x\right)$ yields the final expression for the leading
density gradient correction:
\begin{equation}
P_{1}'\left(x\right)=-\frac{ab\D P}{3^{1/2}2^{1/3}}\left(\frac{2^{5/3}}{\left(x\left(1-x\right)\right)^{1/6}}+I\left(x\right)\right).\label{eq:P1'(x)-1}
\end{equation}
Collecting these results into the ansatz in Eq. (\ref{eq:P' and J ansatz 5/3})
gives the two leading contributions to the anomalous current and density
profile of the L{\'e}vy walk of order $5/3$'' 
\begin{equation}
\begin{cases}
P'\left(x\right)\approx a\D P\biggl[\left(x\left(1-x\right)\right)^{-1/6}-\frac{b}{2^{1/3}3^{1/2}}\\
\text{ }\times\e^{1/3}\left(2^{5/3}\left(x\left(1-x\right)\right)^{-1/6}+I\left(x\right)\right)\biggl]\\
J\approx-\frac{a}{b}v\D P\e^{2/3}\left(1-\frac{2^{4/3}b}{3^{1/2}}\e^{1/3}\right)
\end{cases}.\label{eq:P' and J}
\end{equation}
Equations (\ref{eq:J1}) and (\ref{eq:P1'(x)-1}) are the first finite-size
corrections computed in the context of anomalous transport. To verify
that $P_{1}'\left(x\right)$ and $\mathcal{J}_{1}$ indeed describe
the leading correction to the asymptotic results in Eq. (\ref{eq:J0 P0}),
they are compared to the numerical solutions of the exact L{\'e}vy walk
model equations. These are Eq. (\ref{eq:Derrida's current}) for $J_{exact}$,
Eq. (\ref{eq:Q of x}) for $Q\left(x\right)$ and Eq. (\ref{eq:Q P relation})
which relates $Q\left(x\right)$ to $P\left(x\right)$ as $P\left(x\right)=\frac{\b t_{0}}{\b-1}Q\left(x\right)+\ord{\e^{\b-1}}$.
In Ref . \cite{dhar2013exact} the exact self-consistent equations
for $P\left(x\right)$ and $Q\left(x\right)$ were numerically solved
and shown to agree with Eq. (\ref{eq:Q P relation}). Figure \ref{fig:P_1 collapse}
shows $P_{1}\left(x\right)$ alongside the collapse of $\e^{1/3}\left(P_{Num}\left(x\right)-P_{0}\left(x\right)\right)$
for different values of $\e$. $P_{Num}\left(x\right)$ is obtained
by numerically solving Eq. (\ref{eq:Q of x}) for $Q_{Num}\left(x\right)$
and then using Eq. (\ref{eq:Q P relation}) to relate $Q_{Num}\left(x\right)$
to $P_{Num}\left(x\right)$. Notice that the matching to $P_{1}\left(x\right)$
breaks down near the endpoints. Indeed, the derivation of the approximate
relation between $J$ and $P\left(x\right)$ in Eq. (\ref{eq:Fredholm 2nd})
is valid only for $x\in\left[\e,1-\e\right]$ and, consequently, so
is its solution (see Appendix B). Specifically, the behavior of $P\left(x\right)$
in the intervals $x\in\co{0,\e}\cup\oc{1-\e,1}$ unfortunately remains
out of reach. The same difficulties were reported in Ref. \cite{buldyrev2001average},
which studies the closely related problem of computing the mean first-passage
time for the L{\'e}vy walk. Nevertheless, it is straightforward to show
that limiting the domain of $x$ to $\left[\e,1-\e\right]$ does not
introduce new corrections. Figure \ref{fig:J} compares $J$ of Eq.
(\ref{eq:P' and J}) to the asymptotic current $J_{0}=\e^{2/3}\mathcal{J}_{0}$
and to the exact current $J_{exact}$, obtained by numerically solving
Eq. (\ref{eq:Q of x}) for $Q\left(x\right)$ and substituting the
result into Eq. (\ref{eq:Derrida's current}) for $J_{exact}$.

\section{Other $\protect\b$ Values and Higher Order Corrections \label{sec:Different beta}}

Let us finally discuss the application of this method to other values
of$\b$ and higher-order corrections. For a general $\b$ and arbitrary
order, this method faces two caveats: The first is due to the fact
that Eq. (\ref{eq:Fredholm 2nd}) for $J$ is a small-$\e$ approximation
of Eq. (\ref{eq:Derrida's current}) for $J_{exact}$, implying that
some higher-order corrections must have been neglected in its derivation.
The second follows from limitations on the source term\textquoteright s
behavior at the boundaries that are imposed by the Sonin formula.
These two caveats are explained next and additional details are provided
in Appendix B.

The appropriate ansatz for $P'\left(x\right)$ and $J$ for a L{\'e}vy
walk of order $\b$ is
\begin{equation}
\begin{cases}
P'\left(x\right)=\sum_{m=0}^{M}\e^{\left(2-\b\right)m}P_{m}'\left(x\right)+\ord{\e^{\left(2-\b\right)\left(M+1\right)}}\\
J=\e^{\b-1}\left[\sum_{m=0}^{M}\e^{\left(2-\b\right)m}\mathcal{J}_{m}+\ord{\e^{\left(2-\b\right)\left(M+1\right)}}\right]
\end{cases},\label{eq:General beta ansatz}
\end{equation}
where $M$ is the maximal expansion order beyond which the method
is no longer accurate. $M$ is the manifestation of the first caveat
mentioned above. From Eqs. (\ref{eq:General beta ansatz}) and (\ref{eq:first few equations}),
one learns that the hierarchy equation for $P_{M}'\left(x\right)$
is of $\ord{\e^{\b-1+\left(2-\b\right)M}}$. Thus, to determine $M$
we must account for the terms neglected in the derivation of Eq. (\ref{eq:Fredholm 2nd})
for $J$ (see Appendix A) and find the order at which they enter the
equation for $P_{M}'\left(x\right)$. Appendix A shows that the leading
$\ord{\e^{2\left(\b-1\right)}}$ corrections in Eq. (\ref{eq:Fredholm 2nd})
set $M=\left\lceil \frac{\b-1}{2-\b}\right\rceil $. 

It is important to stress that the hierarchy equations for $P'_{m}\left(x\right)$
are perfectly accurate for $0\le m<M$. Moreover, if only the anomalous
current $J$ is of interest, one can significantly increase $M$ by
working directly with Eq. (\ref{eq:J of Q}). Then the $\ord{\e^{2\left(\b-1\right)}}$
corrections are replaced by $\ord{\e^{3}}$ corrections and $M$ increases
to $M=\left\lceil \frac{4-\b}{2-\b}\right\rceil $. 

The second caveat is intrinsic to the Sonin formula. As explained
in Ref, \cite{samko1993fractional}, the Sonin formula applies only
when the source term's boundary singularity is weaker than the kernel's
singularity. Depending on the value of $\b$, some of the hierarchy
equations might not satisfy this requirement, even for $m<M$. Equation
(\ref{eq:first few equations}) shows that the source term in the
hierarchy equation for $P_{m}'\left(x\right)$ is proportional to
$P_{m-1}'\left(x\right)$. In Appendix B, the boundary behavior of
$P_{m-1}'\left(x\right)$ is argued to be of the form $P_{m-1}'\left(x\ra0\right)\pro x^{\left(2m-1\right)\left(\frac{\b-2}{2}\right)}$
for general $m$ and $\b$, with similar behavior for $x\ra1$. Comparing
this singularity to the kernel's singularity $\b-1$ restricts the
Sonin formula to $m<\frac{\b}{2\left(2-\b\right)}$. 

Nevertheless, since all hierarchy equations satisfying $0\le m<M$
are precise, hierarchy equations for $P_{m}'\left(x\right)$ with
$\frac{\b}{2\left(2-\b\right)}<m<M$ can still be solved by any other
method, be it analytical or numerical, and yield the correction solutions.

\section{Conclusions \label{sec:CONCLUSIONS}}

In this paper, the anomalous transport properties of a 1D L{\'e}vy walk
of order $\b$ are studied on a finite interval of size $L$ under
nonequilibrium settings. Extending the work of Ref. \cite{dhar2013exact},
which related the anomalous walker current $J$ to the density gradient
$P'\left(x\right)$ for asymptotically large $L$, a more general
integral equation which also captures finite-size corrections is derived.
A perturbative method is presented for constructing an order-by-order
solution of this equation. The method is explicitly demonstrated by
computing the leading correction to the asymptotic behavior for $\b=5/3$,
and its results are shown to be in excellent agreement with the numerical
solution of the exact equations.

Remarkably, many other physical problems are described by similar
integral equations \cite{kirkwood1948intrinsic,douglas1989surface,lazopoulos2006non,carpinteri2011fractional,Miron2019},
bringing hope that the method presented here could be used in a verity
of different fields. In the context of anomalous transport, it is
interesting to compare the results computed here to simulations and
experiments. This could test if L{\'e}vy walks are indeed a reliable model
for anomalous transport, even beyond the asymptotic limit. Applying
this method to study additional L{\'e}vy walk properties, as well as other
physical problems, is an equally interesting and exciting prospect. 

\section{Acknowledgments \label{sec:ACKNOWLEDGMENTS}}

I would like to thank my advisor D. Mukamel for his ongoing encouragement,
help, and support. I also thank O. Raz and V. V. Prasad for critical
reading of this manuscript as well as G. Falkovich for helpful comments.
In addition, previous projects with Anupam Kundu and Julien Cividini
have significantly influenced this study, and their collaboration
is greatly appreciated. This work was supported by a research grant
from the Center of Scientific Excellence at the Weizmann Institute
of Science.

\section*{Appendix A - The Derivation of Eq. (\ref{eq:Fredholm 2nd}) }

Equation (\ref{eq:Derrida's current}) for $J_{exact}$ was derived
in Ref. \cite{dhar2013exact} and serves as the basis for the derivation
of Eq. (\ref{eq:Fredholm 2nd}) for $J$ and mainly differs in the
treatment of the finite-size corrections. The key steps of the derivation
are outlined next. 

Using $\psi\left(t\right)$ of Eq. (\ref{eq:psi beta}), the second
line of Eq. (\ref{eq:Derrida's current}) for $J_{exact}$ becomes
\[
\frac{L}{2}\int_{0}^{1}\dif y\text{ }\sign{x-y}\p\left(\frac{L\left|x-y\right|}{v}\right)Q\left(y\right)
\]
\[
=\frac{L}{2}\left(\int_{x-\e}^{x}\dif y\text{ }Q\left(y\right)-\int_{x}^{x+\e}\dif y\text{ }Q\left(y\right)\right)
\]
\begin{equation}
+\frac{L\e^{\b}}{2}\left(\int_{0}^{x-\e}\frac{\dif yQ\left(y\right)}{\left(x-y\right)^{\b}}-\int_{x+\e}^{1}\frac{\dif yQ\left(y\right)}{\left(y-x\right)^{\b}}\right).\label{eq:J der first step}
\end{equation}
Expanding the second line of Eq. (\ref{eq:J der first step}) in small
$\e$ yields 
\begin{equation}
-\frac{t_{0}v\e}{2}Q'\left(x\right)+\ord{\e^{3}}\label{eq:first line}
\end{equation}
and integrating the third line by parts yields 
\[
\frac{t_{0}v\e}{2-\b}Q'\left(x\right)+\frac{t_{0}v\e^{\b-1}}{2\left(\b-1\right)}
\]
\begin{equation}
\times\left(\frac{Q\left(1\right)}{\left(1-x\right)^{\b-1}}-\frac{Q\left(0\right)}{x^{\b-1}}-\int_{0}^{1}\frac{\dif yQ'\left(y\right)}{\left|x-y\right|^{\b-1}}\right)+\ord{\e^{3}},\label{eq:second line}
\end{equation}
where $Q\left(0\right)=Q_{L}$ and $Q\left(1\right)=Q_{R}$ follow
from Eq. (\ref{eq:baths}). Collecting these terms back into $J_{exact}$
gives
\begin{equation}
J=\frac{t_{0}v\b\e Q'\left(x\right)}{2\left(2-\b\right)}-\frac{t_{0}v\e^{\b-1}}{2\left(\b-1\right)}\int_{0}^{1}\frac{\dif yQ'\left(y\right)}{\left|x-y\right|^{\b-1}}+\ord{\e^{3}}.\label{eq:J of Q}
\end{equation}
To obtain Eq. (\ref{eq:Fredholm 2nd}), which relates $J$ and $P'\left(x\right)$,
one uses the relation $P\left(x\right)=\frac{\b t_{0}}{\b-1}Q\left(x\right)+\ord{\e^{\b-1}}$
in Eq. (\ref{eq:Q P relation}), which inevitably introduces corrections
of $\ord{\e^{2\left(\b-1\right)}}$ into Eq. (\ref{eq:J of Q}). 

Two important comments are in order: The first is that Eq. (\ref{eq:Fredholm 2nd})
is valid only for $\b>\frac{3}{2}$ due to the neglected $\ord{\e^{2\left(\b-1\right)}}$
corrections. The second is that the manipulations involved in going
from Eq. (\ref{eq:J der first step}) to Eq. (\ref{eq:Fredholm 2nd})
are valid only for $x\in\left[\e,1-\e\right]$. This implies that
the density profiles in Eq. (\ref{eq:P' and J}) are not accurate
for $x\in\co{0,\e}\cup\oc{1-\e,1}$. 

\section*{Appendix B - The Sonin Formula and Its Solubility Condition }

\subsubsection{The Sonin Formula}

The Sonin formula provides the formal solution to a class of WSFIEs
of the first kind. Specifically, it can be used to solve equations
of the form 
\begin{equation}
h\left(x\right)=\int_{0}^{1}\dif y\frac{\vf\left(y\right)}{\left|x-y\right|^{\b-1}}\label{eq:A-S-Fredholm first kind}
\end{equation}
for $\vf\left(x\right)$ where $1<\b<2$. For the purpose of this
study, it is sufficient to only consider source terms $h\left(x\right)$
which are symmetric under reflections $x\ra1-x$ and are of the form
\begin{equation}
h\left(x\right)=\frac{h^{*}\left(x\right)}{\left(x\left(1-x\right)\right)^{\left(\b-1\right)-\g}},\label{eq:A-S-h}
\end{equation}
where $h^{*}\left(x\right)$ a smooth function of $x$ and $0<\g<\b-1$.
The latter condition means that the Sonin formula applies only when
the source term\textquoteright s boundary singularity is weaker than
the kernel\textquoteright s singularity. For $\g$ and $h\left(x\right)$
satisfying these conditions, the Sonin formula yields the solution
\begin{equation}
\vf\left(x\right)=\frac{\mathcal{B}}{x^{\frac{2-\b}{2}}}\der{}x\int_{x}^{1}\frac{\dif tt^{2-\b}}{\left(t-x\right)^{\frac{2-\b}{2}}}\der{}t\int_{0}^{t}\frac{\dif qq^{\frac{\b-2}{2}}h\left(q\right)}{\left(t-q\right)^{\frac{2-\b}{2}}}\label{eq:A-S-Sonin sol}
\end{equation}
where $\mathcal{B}=-\frac{\sin\left[\frac{\pi\b}{2}\right]\G\left[\b-1\right]}{\pi\G\left[\frac{\b}{2}\right]^{2}}$
and $\G\left[x\right]$ is the gamma function. It is important to
stress that the Sonin formula applies to far more general WSFIE's
of the first kind. An extensive account and further details can be
found in Refs. \cite{samko1993fractional,buldyrev2001average,dhar2013exact}.

\subsubsection{Solubility Condition}

Next, the implications of the requirements on the boundary singularity
of $h\left(x\right)$ of Eq. (\ref{eq:A-S-h}) are discussed. Consider
the hierarchy of integral equations obtained by substituting the ansatz
of Eq. (\ref{eq:General beta ansatz}) into Eq. (\ref{eq:Fredholm 2nd}).
The first equation is
\begin{equation}
-\frac{2\b\mathcal{J}_{0}}{v}=\int_{0}^{1}\dif y\frac{P_{0}'\left(y\right)}{\left|x-y\right|^{\b-1}}\label{eq:first eqn}
\end{equation}
and its constant source term trivially satisfies the requirements
in Eq. (\ref{eq:A-S-h}). Imposing the boundary conditions in Eq.
(\ref{eq:baths}) provides the asymptotic solution 
\begin{equation}
P_{0}'\left(x\right)=\frac{\G\left[\b\right]\D P}{\G\left[\frac{\b}{2}\right]^{2}\left(x\left(1-x\right)\right)^{\frac{2-\b}{2}}}.\label{eq:first eqn solution}
\end{equation}

The next equation, now for the leading correction $P_{1}'\left(x\right)$,
is
\begin{equation}
\b\left(\frac{\b-1}{2-\b}P_{0}'\left(x\right)-\frac{2}{v}\mathcal{J}_{1}\right)=\int_{0}^{1}\dif y\frac{P_{1}'\left(y\right)}{\left|x-y\right|^{\b-1}}.\label{eq:A-S-P_1'(x) Eq}
\end{equation}
The only nonconstant source term in this equation is $\pro P_{0}'\left(x\right)$.
The range of $\b$ for which its singularity is weaker than that of
the kernel is 
\begin{equation}
\b>\frac{4}{3}.\label{eq:A-S- P_1'(x) beta range}
\end{equation}
As such, the leading correction $P_{1}'\left(x\right)$ can be computed
from the Sonin formula for any $\b$ in this range. 

The general equation for $P_{m}'\left(x\right)$,
\begin{equation}
c_{1}P_{m-1}'\left(x\right)-c_{2}\mathcal{J}_{m}=\int_{0}^{1}\dif y\frac{P_{m}'\left(y\right)}{\left|x-y\right|^{\b-1}},\label{eq:general m eqn}
\end{equation}
with $c_{1}=\frac{2\b}{v}\frac{v\left(\b-1\right)}{2\left(2-\b\right)}$
and $c_{2}=\frac{2\b}{v}$, is used next to find the range of allowed
$\b$ at any order $m$. To this end, let us take the leading singular
behavior of the source term $P_{m-1}'\left(x\right)$ to be $\pro\left(x\left(1-x\right)\right)^{-\g}$.
For $P_{m-1}'\left(x\right)$ to satisfy the requirements of Eq. (\ref{eq:A-S-h}),
$\g$ can only take values in $0<\g<\b-1$. The solution of this equation
via the Sonin formula is
\begin{equation}
P_{m}'\left(x\right)\pro-x^{\frac{\b-2}{2}}\der{}x\int_{x}^{1}\dif t\frac{t^{2-\b}Y\left(t\right)}{\left(t-x\right)^{\frac{2-\b}{2}}}\label{eq:A-S-P_m'(x)}
\end{equation}
where 
\begin{equation}
Y\left(t\right)=\der{}t\int_{0}^{t}\dif q\frac{\left(q\left(1-q\right)\right)^{-\g}}{\left(q\left(t-q\right)\right)^{\frac{2-\b}{2}}}\label{eq:Y(t)}
\end{equation}
and the term $\pro\mathcal{J}_{m-1}$ was neglected since its boundary
singularity is trivially weaker than that of $\left(x\left(1-x\right)\right)^{-\g}$. 

To continue, note that, although not manifest in Eq. (\ref{eq:Fredholm 2nd}),
the hierarchy ansatz reveals the symmetry of $P'\left(x\right)$ under
reflections $x\ra1-x$. To see this, note that the source term in
Eq. (\ref{eq:first eqn}) for $P_{0}'\left(x\right)$ is independent
of $x$. It is easy to show that this equation is symmetric under
$x\ra1-x$ and so is its solution. Next, since $P_{0}'\left(x\right)$
is the only nonconstant source term in Eq. (\ref{eq:A-S-P_1'(x) Eq})
for $P_{1}'\left(x\right)$, one can show that $P_{1}'\left(x\right)$
must too be symmetric under inversion. Using induction one can show
this symmetry propagates throughout the entire hierarchy. It is thus
sufficient to consider the behavior of $P_{m}'\left(x\right)$ for
$x\ra0$. One can then use Eq. (\ref{eq:A-S-P_m'(x)}) to show that
the leading boundary singularity of $P_{m}'\left(x\right)$ is 
\begin{equation}
P_{m}'\left(x\ra0\right)\pro x^{\b-2-\g}.\label{eq:S' singularity}
\end{equation}

By comparing Eq. (\ref{eq:S' singularity}) to the boundary singularity
for the first few values of $m$, the range of allowed $\b$ for any
order $m$ can be obtained: The boundary singularity of $P_{1}'\left(x\right)$,
whose source term is $\pro\left(x\left(1-x\right)\right)^{\frac{\b-2}{2}}$,
is found by setting $\g=\frac{2-\b}{2}$ and yields
\begin{equation}
P_{1}'\left(x\ra0\right)\pro x^{-3\left(\frac{2-\b}{2}\right)}.\label{eq:P_1'(x)}
\end{equation}
Next, the boundary singularity of $P_{2}'\left(x\right)$, whose source
term is $\pro\left(x\left(1-x\right)\right)^{3\left(\frac{\b-2}{2}\right)}$,
is found by setting $\g=3\left(\frac{2-\b}{2}\right)$ and yields
\begin{equation}
P_{2}'\left(x\ra0\right)\pro x^{-5\left(\frac{2-\b}{2}\right)}.\label{eq:P_2''(x)}
\end{equation}
Repeating this process, one finds that the boundary singularity for
general $m$ is 
\begin{equation}
P_{m}'\left(x\ra0\right)\pro x^{-\left(2m+1\right)\left(\frac{2-\b}{2}\right)}.\label{eq:P_m''(x)}
\end{equation}
As such, the highest order correction $P_{m}'\left(x\right)$ which
can be computed by the Sonin formula, for a given $\b$, is obtained
by comparing the singularity of the source term $P_{m-1}'\left(x\ra0\right)\pro x^{-\left(2m-1\right)\left(\frac{2-\b}{2}\right)}$
in the equation for $P_{m}'\left(x\right)$ to the kernel singularity,
providing the relation
\begin{equation}
m<\frac{\b}{2\left(2-\b\right)}.\label{eq:A-S-order m}
\end{equation}

\bibliographystyle{unsrt}
\bibliography{bib}

\end{document}